\documentclass[pre,preprint,aps,nofootinbib,showpacs]{revtex4}
\usepackage{amssymb}

\usepackage{graphics,graphicx,epsfig}
\usepackage{amsmath}
\usepackage{graphicx}

\newcommand{\be}{\begin{equation}}
\newcommand{\ee}{\end{equation}}
\newcommand{\ba}{\begin{eqnarray}}
\newcommand{\ea}{\end{eqnarray}}

\begin{document}

\title{Semi-classical Scar functions in phase space}
\author{Alejandro M.F Rivas \footnote{member of the CONICET}}
\date{\today {}}

\begin{abstract}
We develop a semi-classical approximation for the scar function in the
Weyl-Wigner representation in the neighborhood of a classically unstable
periodic orbit of chaotic two dimensional systems. The prediction of hyperbolic
fringes, asymptotic to the stable and unstable manifolds, is verified
computationally for a (linear) cat map, after the theory is adapted to a
discrete phase space appropriate to a quantized torus. 
Characteristic fringe patterns can be distinguished even for quasi-energies 
where the fixed point is not Bohr-quantized. Also the patterns are highly
localized in the neighborhood of the periodic orbit and along its stable and
unstable manifolds without any long distance patterns that appears for the case
of the spectral Wigner function.
\end{abstract}

\affiliation{Departamento de F\'{\i}sica, Comisi\'on Nacional de Energ\'{\i}a At\'omica. Av. del Libertador 8250, 1429 Buenos Aires, Argentina. \\
and\\
Instituto de Ciencias, Universidad Nacional de General Sarmiento,
J.M. Gutierrez 1150, 1613 Los Polvorines Prov. Buenos Aires, Argentina.\\}



\maketitle

\noindent 
%

\section{Introduction}

 The Gutzwiller
trace formula provides a tool for the semi-classical evaluation of the energy
spectrum in terms of the periodic orbits of the system. However the number
of long periodic orbits required to resolve the spectrum
increases exponentially with the Heisenberg time $T_{H}$ \cite{gutz}.

%

The semi-classical theory of short periodic orbits developed by Vergini and
co-workers \cite{vergiscars}-\cite{edudavid} is a formalism that allows to obtain all the 
quantum information
of a chaotic Hamiltonian system in terms of a very small number of short
periodic orbits.  In this context, the scar functions play a crucial role.
These wavefunctions, that have been introduced in several previous
works \cite{edudavid}-\cite{faure}, live in the neighborhood of the classical
trajectories, resembling the hyperbolic structure of the phase space in their immediate
vicinity. This property makes them extremely suitable for investigating
chaotic eigenfunctions. Recently it has been shown
that the matrix elements between scar functions provide information about
the heteroclinic classical structure \cite{edudavid,hetero}.

In order to perform further developments to this semi-classical theory 
of short periodic orbits, it is important to provide a general expression 
in phase space for the scar functions 
explicitly in terms of the classical invariants that generates the dynamics
of the system. This is the purpose of this paper. In this context the choice of
the Weyl-Wigner representation  provides a phase space vision
which best allows a quantum classical comparison \cite{truchini}. 

The semiclassical expression here deduced predicts characteristic hyperbolic patterns
located in the neighborhood of the periodic orbits and along its stable
 and unstable manifolds. 

In spite to compare the general expression here founded with a "realistic"
system the cat map is employed , i.e. 
the quantization of linear symplectic maps on the torus. 
As was shown by Keating 
\cite{keat2} in this case the semiclassical theory is exact, making this maps an ideal probe
for our expressions.  After the formalism is adapted for a torus phase space, 
we  see an important agreement between the semi-classical construction of the and the
numerically computed Wigner scar functions for the cat maps. The characteristic hyperbolic patterns are clearly discernible even for non Bohr-quantized values of the classical action.

Although most studies in phase space have adopted the Hussimi representation
\cite{vornon,faure} this must be interpreted as a Gaussian Smoothing of the Wigner
function in an area of size $\hbar$ that usually dampens the fine structure
extending along the stable and unstable manifolds \cite{truchini}.

This work is a first step to  further objectives that includes the
investigation of the matrix elements between scar functions.


Section 2 deals with the definition of scar functions and develops its 
relationship with the spectral operator. Then, the Weyl-Wigner representation 
of the scar function, here called Scar Wigner Function, is studied. 
Mainly, its local hyperbolic form of the neighborhood of a single periodic orbit 
 is obtained only in terms of classical objects.
 Also, this expression is compared to the already studied spectral Wigner function \cite{wigscar} .

Section 3 is devoted to study the particular case of the cat map where not only the
semi-classical theory is exact but also the linear approximation is valid
throughout the torus. The scar Wigner function are then clearly visualized 
as a hyperbolic fringe pattern that agree with the semiclassical expression derived 
in section 2.
It has to be noted that we here choose values of N such that the scar states
are not eigenfunctions of the cat map as Faure et al. \cite{faure} shows it
happens. Also the fringe patterns  here analyzed lie entirely within the
 scope of standard semiclassical theory, restricted to fairly short times.
The interesting question concerning homoclinic recurrences depends on dynamics 
for longer times.

\section{Scar Wigner functions}

Scar function states are the main object of study of the current work.
According to \cite{edudavid}-\cite{faure}, the scar function $\left| \varphi
_{X,\phi}\right\rangle $ of parameter $\phi$
 constructed on a single periodic point $X=\left( P,Q\right) $ is defined as 
\begin{equation}
\left| \varphi_{X,\phi}\right\rangle =\int_{-\frac{T}{2}}^{\frac{T}{2}}dte^{i\phi
t}\cos\left( \frac{\pi t}{T}\right) \widehat{U}^{t}\left| X\right\rangle
\label{scarfun}
\end{equation}
where $T=\ln\hbar$ is the Ehrenfest time, $\left| X\right\rangle $ is a
coherent state centered in the point $X$ on the periodic
orbit and $\widehat{U}^{t}$ is the unitary propagator that governs the
quantum evolution of the system. This wavefunctions have been shown, in the
Hussimi representation, to live in the neighborhood of the trajectory,
resembling the hyperbolic structure of the phase space in their immediate
vicinity \cite{faure}.

On the other hand  the spectral
operator (or energy Green function) is defined as
\begin{equation}
\widehat{G}_{E,\varepsilon}=\int_{-\infty}^{\infty}dte^{i\phi t}\widehat
{U}^{t}e^{-\frac{t\left| \varepsilon\right| }{\hbar}}  \label{spec}
\end{equation}
with $\widehat{U}^{t}=e^{\frac{-i}{\hbar}\widehat{H}t}$ and $\phi=\frac
{E}{\hbar}$ so that 
\begin{align}
\widehat{G}_{E,\varepsilon} & =\frac{1}{\pi\hbar}\int_{-\infty}^{\infty
}dte^{\frac{i}{\hbar}\left( E-\widehat{H}\right) t}e^{-\frac{t\left|
\varepsilon\right| }{\hbar}}=-\frac{1}{\pi}\Im\left( \frac {1}{%
\widehat{H}-E-i\varepsilon}\right) \\
& =\frac{1}{\pi}\frac{\varepsilon}{\left( \widehat{H}-E\right)
^{2}+\varepsilon^{2}}\equiv\delta_{\varepsilon}\left( E-\widehat{H}\right)
=\sum_{n}\left| \psi_{n}\right\rangle \left\langle \psi_{n}\right|
\delta_{\varepsilon}\left( E-E_{n}\right)
\end{align}
where $\left| \psi_{n}\right\rangle $ $\ $are the Hamiltonian
eigenfunctions with energy $E_{n}$ and $\delta_{\varepsilon}\left( x\right) 
$ is a normalized function whose width $\varepsilon$ can be taken to be
arbitrary small, so that $\delta_{\varepsilon}$ tend to the Dirac $\delta$
function as $\varepsilon\rightarrow0.$ Then $\widehat{G}_{E,\varepsilon}$
is not a pure state but a statistical mixture. In the limit $%
\varepsilon\rightarrow0$ the spectral operator is a comb of delta functions
on the eigenangles, whose amplitudes are the corresponding individual
density matrices $|\psi_{n}><\psi_{n}|$. For values of $\varepsilon$ larger
than the mean level spacing, several eigenstates contribute to the spectral
operator in a Lorentzian -like smoothing of energy width $\varepsilon$.

It is important to mention here that the cosine term in (\ref{scarfun}) has
been chosen so that the energy spreading is minimum \cite{edudavid} but can
be replaced by any damping term with characteristic time $T$ as for example, $%
e^{-\frac{t\left| \varepsilon\right| }{\hbar}}$ with $\varepsilon
\thickapprox\hbar/T$. So that, it is possible to establish a relationship of the
 scar functions with the spectral operator
\begin{equation}
|\varphi_{X,\phi}>\cong\widehat{G}_{\hbar\phi,\hbar/T}|X>
\end{equation}
that is, the scar function is the spectral operator acting on the coherent state
centered at the point $X$ on the periodic orbit.

The purpose of this work, is to study the scar function in phase space by means of the
Weyl Wigner representation $\rho_{_{X,\phi}}(x)$, here call the Scar Wigner
function, 
\begin{equation}
\rho_{X,\phi}(x)=tr\left[ \widehat{R}_{x}\widehat{\rho}_{X,\phi}\right]
\label{SWF}
\end{equation}
where $\widehat{\rho}_{X,\phi}\equiv\left| \varphi_{X,\phi}\right\rangle \left\langle
\varphi_{X,\phi}\right| $, is the density matrix of the scar function and $%
\widehat{R}_{x}$ are the set of reflection operators thought  points $%
x=\left( p,q\right) $ in phase space \cite{ozrep,opetor}. The Weyl-Wigner representation, 
defined through the set of reflection operators (see Appendix),
has the advantage to show structures of size lower than $\hbar$ \cite{zurek}
while the the Hussimi representation  is a Gaussian smoothing
on a region of size $\hbar$ \cite{glauber,truchini} and hence has a lower resolution.

For the case of the spectral operator its Wigner function 
\begin{equation}
G_{E,\varepsilon}(x)=W(E,\varepsilon,x)=tr\left[ \widehat{R}_{x}\widehat{%
G}_{E,\varepsilon}\right]
\end{equation}
is known as the Spectral Wigner function, first introduced by Berry \cite{berry}
, has been recently shown to present important scarring features \cite
{wigscar} some of which would be present in $\rho_{_{X,\phi}}(x)$ as will be  seen
in this work.

It must be noted that by construction 
\be
\widehat{\rho }_{X,\phi}\equiv |\varphi
_{X,\phi}><\varphi _{X,\phi}|  \label{rhoscar}
\ee
 is a pure state while $\widehat{G }_{E,\varepsilon }$
is a statistical average but both are related by 
\[
\widehat{\rho }_{X,\phi}\cong 
\widehat{G }_{\hbar\phi,\hbar/T }|X><X|\widehat{G }_{\hbar\phi,\hbar/T
}^{\dagger },
\]
then for the scar Wigner function 
\begin{equation}
\rho _{X,\phi}(x)=tr\left[ \widehat{R}_{x}\widehat{\rho }_{X,\phi}\right] \cong <X|\widehat{%
G}_{\hbar\phi,\hbar/T }^{\dagger }\widehat{R}_{x}\widehat{G }%
_{\hbar\phi,\hbar/T }|X>
\end{equation}

Inserting the definition of scar function (\ref{scarfun})  and  (\ref{rhoscar})
in (\ref{SWF}) it can be seen that 
\begin{equation}
\rho _{X,\phi}(x)=\int_{-\frac{T}{2}}^{\frac{T}{2}}\int_{-\frac{T}{2}}^{\frac{T}{2%
}}dt^{\prime }dte^{i\phi t}\cos \left( \frac{\pi t}{T}\right) e^{-i\phi
t^{\prime }}\cos \left( \frac{\pi t^{\prime }}{T}\right) <X|\widehat{U}%
^{-t^{\prime }}\widehat{R}_{x}\widehat{U}^{t}|X>.
\end{equation}
Let us now use the decomposition of the propagator in terms of reflection
operators \cite{ozrep} 
\begin{equation}
\widehat{U}^{t}=\left( \frac{1}{\pi \hbar }\right) ^{L}\int dxU^{t}(x)%
\widehat{R}_{x}
\end{equation}
where $\int dx$ is an integral over the whole phase space and $L$ is the
number of degrees of freedom. So that, 
\begin{align}
\rho _{X,\phi}(x)& =\left( \frac{1}{\pi \hbar }\right) ^{2L}\int_{-\frac{T}{2}}^{%
\frac{T}{2}}\int_{-\frac{T}{2}}^{\frac{T}{2}}dt^{\prime }dt\exp \left( i\phi
(t-t^{\prime })\right) \cos \left( \frac{\pi t}{T}\right) \cos \left( \frac{%
\pi t^{\prime }}{T}\right)   \notag \\
& \times \int dx_{1}\int dx_{2}U^{-t^{\prime
}}(x_{2})U^{t}(x_{1})\left\langle X\right| \widehat{R}_{x_{2}}\widehat{R}_{x}%
\widehat{R}_{x_{1}}\left| X\right\rangle.   \label{scarcomple}
\end{align}
The coherent states on points $X=\left( P,Q\right) $ in phase space are
obtained by translating to $X$ the ground state of the harmonic oscillator,
its position representation is 
\begin{equation}
\left\langle q\right. \left| X\right\rangle =\left( \frac{m\omega }{\pi
\hbar }\right) ^{\frac{1}{4}}\exp \left[ -\frac{\omega }{2\hbar }(q-Q)^{2}+i%
\frac{P}{\hbar }\left( q-\frac{Q}{2}\right) \right] .  \label{eq:qX}
\end{equation}
For simplicity, unit frequency $\left( \omega =1\right) $ and mass $%
\left( m=1\right) $ are chosen for the harmonic oscillator without loss of generality.
The overlap of two coherent states is then 
\begin{equation}
\left\langle X\right. \left| X^{\prime }\right\rangle =\exp \left[ -\frac{%
\left( X-X^{\prime }\right) ^{2}}{4\hbar }-\frac{i}{2\hbar }X\wedge
X^{\prime }\right] .  \label{cohcoh}
\end{equation}
Where the wedge product
\begin{equation*}
X\wedge X^{\prime }=PQ^{\prime }-QP^{\prime }=\left( {\mathcal{J}}X\right)
.X^{\prime },
\end{equation*}
with the second equation also defines the symplectic matrix ${\mathcal{J.}}$
As is shown in the Appendix the action of the reflection operator $\widehat{R}%
_{x}$ on a coherent state $\left| X\right\rangle $ is the $x$ reflected
coherent state 
\begin{equation}
\widehat{R}_{x}\left| X\right\rangle =\exp \left( \frac{i}{\hbar }X\wedge
x\right) \left| 2x-X\right\rangle   \label{reflexcoh}
\end{equation}
and the product of three reflections also gives a reflection 
\begin{equation}
\widehat{R}_{x_{2}}\widehat{R}_{x}\widehat{R}_{x_{1}}=e^{\frac{i}{\hbar }%
\Delta _{3}(x_{2},x_{1},x)}\widehat{R}_{x_{2}-x+x_{1}}  \label{3reflex}
\end{equation}
where $\Delta _{3}(x_{2},x_{1},x)$ is the area of the oriented triangle
whose sides are centered on the points $x_{2},x_{1}$ and $x$ respectively.

Then applying (\ref{3reflex}) then (\ref{reflexcoh}) and performing the overlap (%
\ref{cohcoh}) it is  shown that
\begin{equation}
\left\langle X\right| \widehat{R}_{x_{2}}\widehat{R}_{x}\widehat{R}%
_{x_{1}}\left| X\right\rangle =e^{\frac{i}{\hbar}\Delta_{3}(x_{2},x_{1},x)}%
\exp-\left[ \frac{\left( X-x_{R}\right) ^{2}}{\hbar}\right] \label{XRRRX}
\end{equation}
with
\be
x_{R}=x_{2}-x+x_{1}.
\ee
Inserting (\ref{XRRRX}) in (\ref{scarcomple}) in order to perform the double phase space integrals 
\begin{align}
I & =\left( \frac{1}{\pi\hbar}\right) ^{2L}\int dx_{1}\int
dx_{2}U^{-t^{\prime}}(x_{2})U^{t}(x_{1})\left\langle X\right| \widehat{R}%
_{x_{2}}\widehat{R}_{x}\widehat{R}_{x_{1}}\left| X\right\rangle  \notag \\
& =\left( \frac{1}{\pi\hbar}\right) ^{2L}\int dx_{1}\int
dx_{2}U^{-t^{\prime}}(x_{2})U^{t}(x_{1})e^{\frac{i}{\hbar}%
\Delta_{3}(x_{2},x_{1},x)}\exp\left[ -\frac{\left( X-x_{R}\right) ^{2}}{\hbar%
}\right] .  \label{inte}
\end{align}
If the exponential term is omitted in (\ref{inte}), the double phase space
integral is simply performed using the product of symbols in phase space 
\cite{ozrep} that is 
\begin{equation}
U^{t-t^{\prime}}(x)=\widehat{U}^{-t^{\prime}}\widehat{U}^{t}(x)=\left( \frac{%
1}{\pi\hbar}\right) ^{2L}\int dx_{1}\int
dx_{2}U^{-t^{\prime}}(x_{2})U^{t}(x_{1})e^{\frac{i}{\hbar}%
\Delta_{3}(x_{2},x_{1},x)}
\end{equation}
but the presence of the exponential term has to be kept into account so that
\begin{equation}
I=U^{t-t^{\prime}}(x)\exp\left[ -\frac{\left( X-x_{R}(x)\right) ^{2}}{\hbar}%
\right]
\end{equation}
where now $x_{R}(x)$ is a point in phase space that only depends only on $x$.
Another approach to obtain this result is to perform in (\ref{inte}) the
semi-classical approximation for the propagator 
\begin{equation}
U_{sc}^{t}(x)=\sum_{\gamma}\frac{e^{i\alpha_{\gamma}^{t}}}{\left| \det
M_{\gamma}^{t}+1\right| ^{\frac{1}{2}}}\exp\left( \frac{i}{\hbar}S_{\gamma
t}(x)\right)  \label{USC}
\end{equation}
where the sum is performed over all the classical orbits $\gamma$ whose
center lies on the point $x$ \cite{ozrep}. $\ $Then $S_{\gamma t}(x)$ is the
classical center generating function of the orbit, $M_{\gamma}^{t}=$ $\frac{%
\partial^{2}S_{\gamma t}(x)}{\partial x^{2}}$ stand for the monodromy matrix
and $\alpha_{j}^{t}$ its Maslov index.

Inserting (\ref{USC})\ in (\ref{inte}), we get
\begin{equation}
I=\sum_{\gamma_{1}}\sum_{\gamma_{2}}e^{i\left(
\alpha_{\gamma1}^{t}+\alpha_{\gamma_{2}}^{-t^{\prime}}\right) }\int
dx_{1}\int dx_{2}\frac {\exp\frac{i}{\hbar}\left[ S_{%
\gamma_{1}}^{t}(x_{1})-S_{\gamma_{2}}^{t^{\prime}}(x_{2})+%
\Delta_{3}(x_{2},x_{1},x)\right] }{\left| \det
M_{\gamma_{2}}^{-t^{\prime}}+1\right| ^{\frac{1}{2}}\left| \det
M_{\gamma_{1}}^{t}+1\right| ^{\frac{1}{2}}}\exp\left[ -\frac{\left(
X-x_{R}\right) ^{2}}{\hbar}\right] 
\end{equation}
Now the integrals are performed using the stationary phase approximation. In
this case the phase $\phi(x_{2},x_{1},x)=S_{\gamma_{1}}^{t}(x_{1})-S_{%
\gamma_{2}}^{t^{\prime}}(x_{2})+\Delta_{3}(x_{2},x_{1},x)$ is stationary
(i.e.)\ 
\begin{align}
\frac{\partial\phi(x_{2},x_{1},x)}{\partial x_{1}} & =0 \\
\frac{\partial\phi(x_{2},x_{1},x)}{\partial x_{2}} & =0
\end{align}
for $x_{2}(x),x_{1}(x)$ such that the canonical transformation generated by
 $S_{\gamma_{1}}^{t}(x_{1})$ is combined with that generated by $%
S_{\gamma_{2}}^{t^{\prime}}(x_{2})$ to result in a new canonical transformation
generated by $S_{t-t^{\prime}}^{\gamma}(x)$. That is, the canonical
transformation corresponding to the trajectory $\gamma^{t-t^{\prime}}$ is obtain as the
composition of the canonical transformation of the orbits $\gamma_{1}^{t}$ and $%
\gamma_{2}^{-t^{\prime}}$ respectively. For clarity this situation is depicted in Fig
1. (see also \cite{ozrep}). 

\begin{figure}[h]
\setlength{\unitlength}{1cm} 
\centerline{\epsfxsize=25cm \epsffile{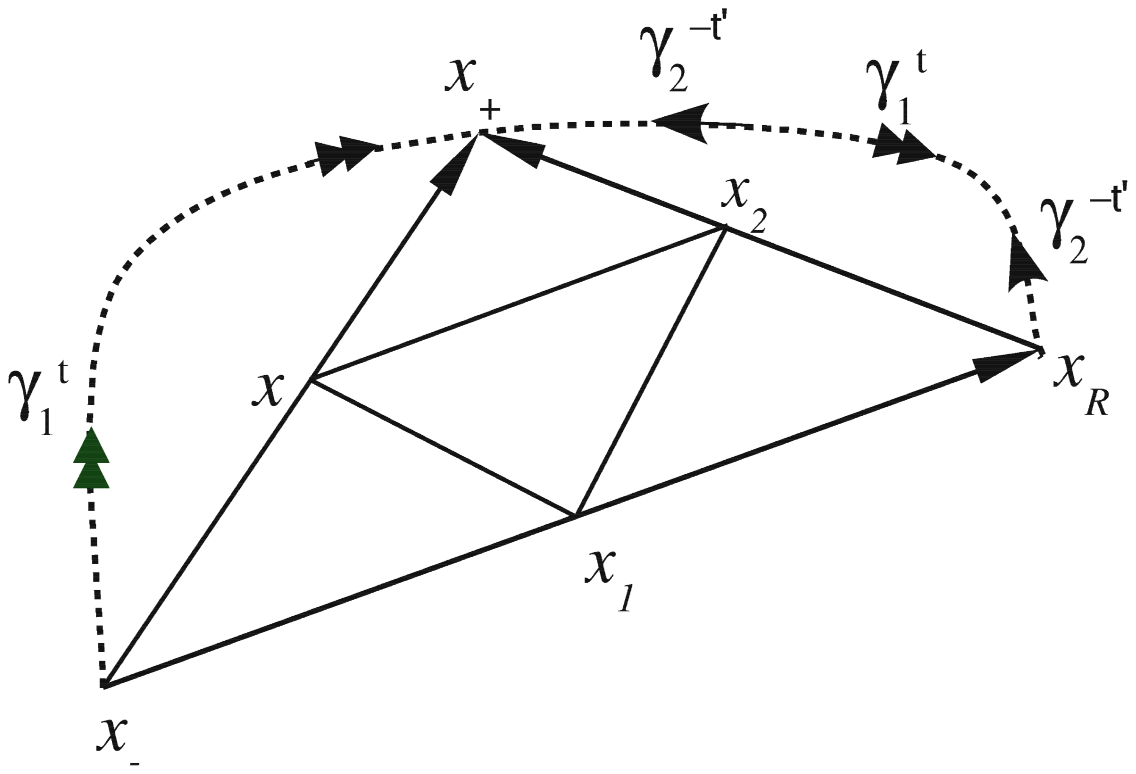} } 
\vspace*{1.0pc}
\caption{{\protect\footnotesize In order to correctly compose the canonical transformations the end point of the orbit $\gamma_1^t$, centered in $x_{1}$   and the initial point of $\gamma_2^{-t^{\prime}}$, centered in $x_{2}$ must coincide in $x_R$. The resulting classical orbit $\gamma^{t-t^{\prime}}$ that joins the points $x_{-}$ and $x_{+}$ and has center point $x$ is the composition of the orbits $\gamma_1^t$  and $\gamma_2^{-t^{\prime}}$.
}}
\label{fig.1}
\end{figure}

Hence 
\begin{equation}
I=\sum_{\gamma}e^{i\alpha_{\gamma}^{t-t^{\prime}}}\frac{\exp\frac{i}{\hbar }%
\left[ S_{\gamma}^{t-t^{\prime}}(x)\right] }{\left| \det M_{\gamma
}^{t-t^{\prime}}+1\right| ^{\frac{1}{2}}}\exp\left[ -\frac{\left(
X-x_{R}(x)\right) ^{2}}{\hbar}\right]  \label{ISC}
\end{equation}
with $x_{R}(x)=x_{2}(x)-x+x_{1}(x)$ corresponding to the end point of the orbit $\gamma_{1}^{t}$
or to the initial point of $\gamma_{2}^{-t^{\prime}}$ as shown in fig. 1.
Denoting $x_{-}$ the initial point of the orbit $\gamma_{1}^{t}$,  its time $t$ evolution is  $x_{R}(x)$  
\begin{equation}
x_{-}(t)=x_{R}(x).
\end{equation}
While $x_{+}$ is the time $t-t^{\prime}$ evolution of $x_{-}$ 
\[
x_{-}(t-t^{\prime})=x_{+},
\]
and $x$ is center point of $x_{-}$  and $x_{+}$ . That is,
\begin{equation}
x=\frac{x_{-}+x_{+}}{2}= \frac{x_{-}+x_{-}(t-t^{\prime})}{2} \label{x}
\end{equation}

The resulting integral (\ref{ISC}) defined in (\ref{inte}) is then inserted in (\ref{scarcomple})
to obtain the semiclassical approximation of the scar Wigner function
\begin{align}
\rho_{X,\phi}^{SC}(x) & =\int_{-\frac{T}{2}}^{\frac{T}{2}}\int_{-\frac{T}{2}}^{%
\frac{T}{2}}dt^{\prime}dte^{i\phi\left( t-t^{\prime}\right) }\cos\left( 
\frac{\pi t}{T}\right) \cos\left( \frac{\pi t^{\prime}}{T}\right)  \notag \\
& \times\sum_{\gamma}e^{i\alpha_{\gamma}^{t-t^{\prime}}}\frac{\exp\frac
{i}{\hbar}\left[ S_{\gamma}^{t-t^{\prime}}(x)\right] }{\left| \det
M_{\gamma}^{t-t^{\prime}}+1\right| ^{\frac{1}{2}}}\exp\left[ -\frac{\left(
X-x_{R}(x)\right) ^{2}}{\hbar}\right]  \label{semicl}
\end{align}

So as to study $\rho_{X,\phi}^{SC}(x)$ in the neighborhood of the periodic point $X$
of the map, the main contribution in the sum over classical orbits in (\ref
{semicl}) will come from the  particular periodic orbit that lies the
 closest from the periodic point $X$. Other orbits contribution will be 
highly damped by the exponential term involving $X-x_{R}(x)$.
Then, only this particular orbit will be taken into account.



For this purpose let us define $x^{\prime}=x-X$. In the same way, $x_{-}$ the
 initial point of the classical orbit $\gamma^{t-t^{\prime}}$ can be written 
as $x_{-}=X+\delta_{-}$ so that its time $t$ evolution  
\begin{equation}
x_{R}=x_{-}(t)=M_{\gamma}^{t}x_{-}=M_{\gamma}^{t}\left( X+\delta_{-}\right) =X+M_{\gamma}^{t}\delta _{-}.
\label{xr}
\end{equation}
Where in the last expressions the flux in  the neighborhood of $X$  has been linearized and  $M_{\gamma}^{t}$ is the symplectic matrix denoting this linearized time evolution. 
Note that for the fixed point $M_{\gamma}^{t}(X)=X$.
Also for the center point defined in (\ref{x})
\begin{equation}
x=X+x^{\prime}=\frac{x_{-}+x_{-}(t-t^{\prime})}{2}=\frac{X+\delta
_{-}+X+M_{\gamma}^{t-t^{\prime}}\delta_{-}}{2}=X+\left( M_{\gamma}^{t-t^{\prime}}+1\right) 
\frac{\delta_{-}}{2}.
\end{equation}
Inverting this last expression 
\begin{equation}
\delta_{-}=2\left( M_{\gamma}^{t-t^{\prime}}+1\right) ^{-1}x^{\prime}
\end{equation}
that is inserted in (\ref{xr}) to obtain 
\begin{equation}
x_{R}=X+2\frac{M_{\gamma}^{t}}{\left( M_{\gamma}^{t-t^{\prime}}+1\right) }x^{\prime}
\end{equation}
and finally,
\begin{equation}
X-x_{R}=-2\frac{M_{\gamma}^{t}}{\left( M_{\gamma}^{t-t^{\prime}}+1\right) }x^{\prime}
 \label{X-xr1}.
\end{equation}
The eigenvalues of the symplectic matrix $M_{\gamma}^{t}$ are $\exp(-\lambda t)$ and $%
\exp(\lambda t)$, ($\lambda$ is the stability exponent of the orbit)
corresponding to the stable and unstable directions generated by the vectors $\vec{\xi_s}$ 
and $\vec{\xi_u}$ respectively. Let us define  $q^{\prime}$ and $p^{\prime}$
as canonical coordinates along the stable and unstable directions respectively 
such that $x^{\prime}=(p^{\prime},q^{\prime}) =q^{\prime}\vec{\xi_s}+p^{\prime}\vec{\xi_u}$,  
with $
\vec{\xi_u}\wedge\vec{\xi_s} =1$.
Then, using (\ref{X-xr1}) with the diagonal representation of the symplectic matrix, the scalar product $\left( X-x_{R}\right) ^{2}=\left( X-x_{R}\right).\left( X-x_{R}\right)$ takes the form
\begin{equation}
\left( X-x_{R}\right) ^{2}=\frac{1}{\cosh{}^{2}\left( \lambda \frac{%
t-t^{\prime}}{2}\right) }\left[ p^{\prime2} e^{ \lambda\left(
t+t^{\prime}\right) } \xi_u^2+ q^{\prime2} e^{-\lambda\left(
t+t^{\prime}\right) } \xi_s^2
 +2 p^{\prime}q^{\prime} \vec{\xi_u} . \vec{\xi_s} \right]  \label{X-xr}
\end{equation}
where $\xi_u^2=\vec{\xi_u}.\vec{\xi_u} $ and $\xi_s^2= \vec{\xi_s}.\vec{\xi_s}$. 
It can also be easily shown that 
\begin{equation}
\left| \det M_{\gamma}^{t}+1\right| ^{\frac{1}{2}}=2\cosh\left( \frac{%
\lambda t}{2}\right).  \label{detM}
\end{equation}

The center action for the orbit that is close to the periodic orbit is
given by 
\begin{equation}
S_{\gamma}^{t}(x)=tS_{X}+x^{\prime}B_{\gamma}^{t}x^{\prime}+O(x^{\prime 3})
\label{action}
\end{equation}
where $S_{X}$ is the action of the periodic orbit (fixed point) for which
the Maslov index $\alpha_{\gamma}^{t}=t\alpha_{\gamma}$. Let us define the
action $\widetilde{S}_{X}=S_{X}+\hbar\alpha_{\gamma}$ in order to include the
Maslov index in the action. $B^{t}$ is the symmetric matrix such that 
\begin{equation}
{\mathcal{J}}B_{\gamma}^{t}=\frac{1-M_{\gamma}^{t}}{1-M_{\gamma}^{t}}
\end{equation}
with 
\begin{equation}
{\mathcal{J}}=\left[ 
\begin{array}{cc}
0 & -1 \\ 
1 & 0
\end{array}
\right]
\end{equation}
Thus $B_{\gamma}^{t}$ is the Cayley parametrization of $M_{\gamma}^{t}$.
Using the stable and unstable directions as coordinate axes, the result is that 
\begin{equation}
{\mathcal{J}}B_{\gamma}^{t}=\left[ 
\begin{array}{cc}
\tanh\left( t\lambda/2\right) & 0 \\ 
0 & -\tanh\left( t\lambda/2\right)
\end{array}
\right] .  \label{cayley}
\end{equation}
Inserting (\ref{cayley}) in (\ref{action}) and putting together with (\ref
{detM}) and (\ref{X-xr}) in (\ref{semicl}) we obtain that the contribution
to the semi-classical scar Wigner function from the orbit that lies close to
the periodic orbit $X$ takes the form 
\begin{align}
\rho_{X,\phi}^{SC}(x) & =\int_{-\frac{T}{2}}^{\frac{T}{2}}\int_{-\frac{T}{2}}^{%
\frac{T}{2}}dt^{\prime}dt\exp i\left\{ \left( \phi+\frac{\widetilde{S}_{X}}{\hbar}
\right) \left( t-t^{\prime}\right) -\frac
{2}{\hbar}p^{\prime}q^{\prime}\tanh\left[ \frac{\left( t-t^{\prime}\right)
\lambda}{2}\right] \right\} \frac{\cos\left( \frac{\pi t}{T}\right) \cos\left( \frac{\pi
t^{\prime}}{T}\right) }{2\cosh\left[ \frac{\left( t-t^{\prime }\right)
\lambda}{2}\right] } \notag \\
\times & \exp\left\{-\frac{1}{\cosh{}^{2}\left( \lambda \frac{%
t-t^{\prime}}{2}\right) }\left[ p^{\prime2} e^{ \lambda\left(
t+t^{\prime}\right) } \xi_u^2+ q^{\prime2} e^{-\lambda\left(
t+t^{\prime}\right) } \xi_s^2
 +2 p^{\prime}q^{\prime} \vec{\xi_u} . \vec{\xi_s}  \right] \right\}
\label{SCSWF}
\end{align}
this last expression shows that the dependence of the scar Wigner function
on the phase space variables has two aspects, a phase oscillating term only
depending on the product $p^{\prime}q^{\prime}$ and a damping term. 
The phase oscillating terms shows phase coherence along the stable and unstable 
directions where the damping factor is null. Phase coherence also holds where the product $%
p^{\prime }q^{\prime}$ is constant, i.e. along each successive hyperbola
that has the stable and unstable manifolds as asymptotes. This implies in
phase oscillations across the hyperbolae, the amplitud of the oscillations decrease with increasing $\lambda$ and will be maximal for $\phi$ corresponding to the Bohr-quantized orbit. 
While the  damping term, implies that away from the asymptotes the amplitude of the phase oscillations presents a Gaussian decreasing with increasing the distance.
 
This amplitude decaying is not present in the Spectral Wigner function that
has strong oscillations away from the periodic orbit and its stable and
unstable manifolds \cite{wigscar}. This facts implies in important superpositions of the
hyperbolic patterns from different orbits that usually washes out all the
structure.


\section{Scar Wigner Functions for the Cat Map}

Now the present theory is applied to the cat map i.e. the linear automorphism
on the $2$-torus generated by the $2\times2$ symplectic matrix $\mathcal{M}$%
, that takes a point $x_{-}$ to a point $x_{+}$ : $x_{+}=\mathcal{M}%
x_{-}\quad\mbox{mod(1)} $. In other words, there exists an integer $2$%
-dimensional vector $\mathbf{m}$ such that $x_{+}=\mathcal{M}x_{-}-\mathbf{m}
$. Equivalently, the map can also be studied in terms of the center
generating function \cite{mcat}. This is defined in terms of center points 
\begin{equation}
x\equiv\frac{x_{+}+x_{-}}2  \label{xdef}
\end{equation}
and chords 
\begin{equation}
\xi\equiv x_{+}-x_{-}=-\mathcal{J}\frac{\partial S(x,\mathbf{m})}{\partial x}%
,  \label{cordef}
\end{equation}
where 
\begin{align}
S(x,\mathbf{m}) & =xBx+x(B-\mathcal{J)}\mathbf{m}+\frac14\mathbf{m}(B+%
\widetilde{\mathcal{J}})\mathbf{m}  \label{sx2}
\end{align}
is the center generating function. Here $B$ is a symmetric matrix (the
Cayley parameterization of $\mathcal{M}$, as in (\ref{cayley})), while 
\begin{equation}
\widetilde{\mathcal{J}}=\left[ 
\begin{array}{c|c}
0 & 1 \\ \hline
1 & 0
\end{array}
\right] .
\end{equation}
We will study here the cat map with the symplectic matrix 
\begin{equation}
\mathcal{M}=\left[ 
\begin{array}{cc}
2 & 3 \\ 
1 & 2
\end{array}
\right] \mbox{, and symmetric matrix } B=\left[ 
\begin{array}{cc}
-\frac{1}{3} & 0 \\ 
0 & 1
\end{array}
\right] .  \label{mhb}
\end{equation}
This map is known to be chaotic, (ergodic and mixing) as all its periodic
orbits are hyperbolic. The periodic points $x_{l}$ of integer period $l$ are
labeled by the winding numbers $\mathbf{m,}$ so that 
\begin{equation}
x_{l}=\left( 
\begin{array}{l}
{p_{l}} \\ 
{q_{l}}
\end{array}
\right) =(\mathcal{M}^{l}-1)^{-1}\mathbf{m} .  \label{xfix}
\end{equation}
The first periodic points of the map are the fixed points at $(0,0)$ and $%
(\frac12, \frac12)$ and the periodic orbits of period 2 are $[(0,\frac12)$ , 
$(\frac12,0)]$, $[(\frac12,\frac16)$ , $(\frac12,\frac56)]$, $[(0,\frac16)$
, $(\frac12,\frac26)]$, $[(0,\frac56)$, $(\frac12,\frac46)]$ and $%
[(0,\frac26), (0,\frac46)]$. The eigenvalues of $\mathcal{M}$ are $%
e^{-\lambda} $ and $e^{\lambda}$ with $\lambda= \ln(2 + \sqrt{3})
\approx1.317 $. This is then the stability exponent for the fixed points,
whereas the exponents must be doubled for orbits of period 2. All the
eigenvectors have directions $\vec{\xi_{s}}=(-\frac{\sqrt{3}}{2},\frac{1}{2})$ 
and $\vec{\xi_{u}}=(1,\frac{1}{\sqrt{3}})$
corresponding to the stable and unstable directions respectively.

Quantum mechanics on the torus, implies a finite Hilbert space of dimension $%
N=\frac{1}{2\pi\hbar}$, and that positions and momenta are defined to have
discrete values in a lattice of separation $\frac{1}{N}$ \cite{hanay,opetor}%
. The cat map was originally quantized by Hannay and Berry \cite{hanay} in
the coordinate representation:

\begin{equation}
<\mathbf{q}_{k}|\hat{\mathbf{U}}_{\mathcal{M}} |\mathbf{q}_{j}> = \left( 
\frac{i}{N} \right) ^{\frac12} {\exp} \left[ \frac{i2\pi}{N}( k^{2} -j k +
j^{2}) \right] ,  \label{uqq}
\end{equation}

where the states $<q|\mathbf{q}_{j}>$ are periodic combs of Dirac delta
distributions at positions $q=j/N mod(1)$, with $j$ integer in $[0,N-1]$. In
the Weyl representation \cite{opetor}, the quantum map has been obtained in 
\cite{mcat} as

\begin{align}
\mathbf{U}_{\mathcal{M}}(x) & =\frac{2}{\left| \det(\mathcal{M} +1)\right|
^{\frac12}} \sum_{\mathbf{m}}e^{i2\pi N\left[ S(x,\mathbf{m})\right] } 
\notag \\
& =\frac{2}{\left| \det(\mathcal{M} +1)\right| ^{\frac12}}\sum_{\mathbf{m }%
}e^{i2\pi N\left[ xBx+x(B-\mathcal{J})\mathbf{m}+\frac14\mathbf{m}(B+%
\widetilde{\mathcal{J}})\mathbf{m}\right] },  \label{ugxp}
\end{align}

where the center points are represented by $x=(\frac{a}{N},\frac{b}{N})$
where $a $ and $b$ are integers in $[0,N-1]$ for odd values of $N$ \cite
{opetor}. There exists an alternative definition of the torus Wigner
function which also holds for even $N$. However, it is constructed on the
quarter torus and this compactification scrambles the hyperbolic patterns.

The fact that the $\mathcal{M}$ matrix has equal diagonal elements implies
in the time reversal symmetry and then the $B $ matrix has no off-diagonal 
elements. This property will be
valid for all the powers of the map and, using (\ref{ugxp}), we can see that
it implies in the quantum symmetry 
\begin{equation}
\mathbf{U}_{\mathcal{M}}^{l}(p,q)= \left( \mathbf{U}_{\mathcal{M}%
}^{l}(-p,q)\right) ^{*} = \left( \mathbf{U}_{\mathcal{M}}^{l}(p,-q)\right)
^{*} .  \label{qsym}
\end{equation}

It has been shown \cite{hanay} that the unitary propagator is periodic
(nilpotent) in the sense that, for any value of $N$ there is an integer $%
k(N) $ such that $\hat{\mathbf{U}}_{\mathcal{M}}^{k(N)}=e^{i\phi} $. Hence
the eigenvalues of the map lie on the $k(N)$ possible sites 
\begin{equation}
\left\{ {\exp} \left[ \frac{i(2m\pi+\phi)}{k(N)} \right] \right\} ,\quad1\le
m\le k(N) .
\end{equation}
For $k(N)< N$ there are degeneracies and the spectrum does not behave as
expected for chaotic quantum systems. In spite of the peculiarities in this
spectra of quantum cat maps, it is likely that non-degenerate states are
typical of chaotic maps, such as very weak nonlinear perturbations of cat
maps that are known to have non degenerate spectra \cite{matos}. Eckhardt 
\cite{Eckhardt} has argued that typically the eigenfunctions of cat maps are
random.

The Scar Wigner Function on the torus depends on the definition of the
periodic coherent state \cite{nonen,sarac2}, with $<p>=P$ and $<q>=Q$. 
In accordance to (\ref{eq:qX})
\begin{equation}
<{\bf X}|\mathbf{q}_{k}>=\sum_{j=-\infty}^{\infty}exp{\ \left\{ 2\pi N\left[ -%
\frac{(j+Q-k/N)^{2}}{2\omega^{2}}-iP(j+\frac{Q}{2}-k/N)\right] \right\} }.
\label{35}
\end{equation}

The Scar function is then defined on the torus as
\begin{equation}
\left| {{\bf \varphi}}_{X,\phi}\right\rangle =\sum_{-T/2}^{T/2}e^{i\phi
t}\cos\left( \frac{\pi t}{T}\right) \mathbf{U}_{\mathcal{M}}^{t}\left| X\right\rangle.
\label{scartoro}
\end{equation}

Note that for maps, time only takes discrete values, then the time integrals in  
(\ref{scarfun}) and  (\ref{SCSWF}) are in this case replaced by summations.
Now, the scar Wigner function on the torus is
\begin{equation}
\mathbf{\rho} _{X,\phi}(x)=\mathbf{Tr}\left[ \widehat{\mathbf{R}}_{x}|{{\bf \varphi}}
_{X,\phi}><{{\bf \varphi}} _{X,\phi}|\right] 
\end{equation}
where the trace is now taken on torus Hilbert space and $\widehat{\mathbf{R}}_{x}$ are the periodic reflection operators on the torus \cite{opetor}.

In order to construct the semiclassical Scar Wigner Functions on the torus we
have to periodized the construction. This is done merely using the recipe \cite{opetor} 
that for any operator its Weyl representation on the torus ${\bf A}(x)$ is obtained from 
is analogue in the plane $A(x)$  by
\begin{eqnarray}
{\bf A}(x) = \sum_{j=-\infty}^{\infty}\sum_{k=-\infty}^{\infty}
(-1)^{2ja+2kb+jkN}A(x+\frac{(k,j)}{2}) ,
\end{eqnarray}
This leads to the property 
\be
{\bf A}(x+\frac{(k,j)}{2}) = (-1)^{2ja+2kb+jkN}{\bf A}(x) ,
\label{w--}
\ee
for the torus Weyl symbol.
In the case of the  Scar Wigner function, the phase factor in (\ref{w--}), leads to four images of the scar pattern, supported by the integer lattice. The images centered on $(P,Q)$, $(P+\frac{1}{2},Q)$ and $(P,Q+\frac{1}{2})$ are all identical, whereas $(P+\frac{1}{2},Q+\frac{1}{2})$ centers a pattern which is the negative of the other ones. This fact has already been studied  for the Wigner function of coherent states \cite{wigscar}.

In figure 2  we compare the exact Scar Wigner function for a cat map with $N=223$, a value for which the quantum map has no degeneracies. In figure 2 (a)  with the semiclassical approximation, correspondingly for  $\hbar= 1/(2\pi N) $, in (d), for the periodic point at $(1/2,1/2)$ whose  action is $ S_X=0.75 $ for a value of $\phi$ that does not Bohr-quantize the orbit. Figure 2.(b) and 2.(c)  show respectively  horizontal and vertical sections of the objects plotted in figures 2.(a) and 2.(d) in the neighborhood of the periodic point. 

It can easily be observed that 
the four images of the scar patters are present both for the exact and semiclassical scar Wigner functions. Also, 
the semiclassical approximation properly describes the overall behavior of the exact dynamical system with detail of oscillations of the order $ 0.02.$

 Although, deviations are present, particularly seen in figure 2.(b), they are due to the contribution to (\ref{SCSWF}) of longer orbits. In particular, the period-2 orbit on the points $[(0,\frac{2}{6})(0,\frac{4}{6})]$ has  $(1/2,1/2)$ as center point.
 This periodic orbits only have contributions for the adequate values of $t-t^{\prime}$, in the specified case $t-t^{\prime}$ must be odd for the period-2 orbit to contribute.

\section{Discussion}

As was already observed for the case of the spectral Wigner function, the imprint of the classical hyperbolicity on the Scar Wigner function is so clear that it can be detected  even for quasi-energies that do not correspond to a Bohr-quantized periodic orbit.

The general features exhibited by our calculations should also be discernible for nonlinear systems as, indeed, our deduction was not restricted to cat maps. 
Though the theory in section II is only local, we conjecture that distorted hyperbolae asymptotic to curved stable and unstable  manifolds  will bear fringes reaching out from the periodic point. In the case of a chaotic Hamiltonian for two degrees of freedom, this pattern should emerge in two-dimensional sections cutting the periodic orbit at a point. This plane should be transverse to that of the orbit,and in the energy shell.

The cut-off with time in our definition of the Scar Wigner function affords equal treatment to all periodic orbits in the denominator of (\ref{SCSWF}). The reason why the main contribution comes from only one hyperbolic orbit is that for an orbit of period $n$ to have coherent contribution in  (\ref{SCSWF}) $t-t^{\prime}$ must have specific values that differ in $n$. That is, there are only $\frac{2T}{n}$ terms instead of $2T$ in (\ref{SCSWF}).
Also, $T$ increases only logarithmically with $\hbar$ so that the semiclassical limit $\lambda \rightarrow n \lambda$ implies in a a cut off for longer orbits.


%

\section{Acknowledgments}

I am  grateful to E. Vergini, D. Wisniacki and D. Schneider for stimulating
discussions and thanks the CONICET for financial support.

%

\section*{Appendix: Reflection Operators in Phase Space}

Among the several representations of quantum mechanics, the Weyl-Wigner
representation is the one that performs a decomposition of the operators
that acts on the Hilbert space, on the basis formed by the set of unitary
reflection operators. In this appendix we review the definition and some
properties of this reflection operators. 

First of all we construct the family of unitary operators 
\begin{equation}
\hat{T}_{q}=\exp (-i\hbar ^{-1}q.\hat{p}),\qquad \hat{T}_{p}=\exp (i\hbar
^{-1}p.\hat{q}),
\end{equation}
and following \cite{ozrep}, we define the operator corresponding to a
general translation in phase space by $\xi =(p,q)$ as 
\begin{eqnarray}
&&\hat{T}_{\xi }\equiv \exp \left( \frac{i}{\hbar }\xi \wedge \hat{x}\right)
\equiv \exp \left[ \frac{i}{\hbar }(p.\hat{q}-q.\hat{p})\right]   \notag \\
&=&\hat{T}_{p}\hat{T}_{q}\ \exp \left[ -\frac{i}{2\hbar }\ p.q\right] =\hat{T%
}_{q}\hat{T}_{p}\ \exp \left[ \frac{i}{2\hbar }\ p.q\right] \ ,
\label{eq:tcor}
\end{eqnarray}
where naturally $\hat{x}=(\hat{p},\hat{q})$. In other words, the order of $%
\hat{T}_{p}$ and $\hat{T}_{q}$ affects only the overall phase of the
product, allowing us to define the translation as above. $\hat{T}_{\xi }$ is
also known as a \textit{Heisenberg operator}. Acting on the Hilbert space we have: 
\begin{equation}
\widehat{T}_{\xi }|q_{a}>=e^{\frac{i}{\hbar }p(q_{a}+\frac{q}{2})}|q_{a}+q>
\label{eq:tq}
\end{equation}
and 
\begin{equation}
\widehat{T}_{\xi }|p_{a}>=e^{-\frac{i}{\hbar }q(p_{a}+\frac{p}{2})}|p_{a}+p>.
\end{equation}
We, hence, verify their interpretation as translation operators in phase
space. The group property is maintained within a phase factor: 
\begin{equation}
\hat{T}_{\xi _{2}}\hat{T}_{\xi _{1}}=\hat{T}_{\xi _{1}+\xi _{2}}\ \exp [%
\frac{-i}{2\hbar }\xi _{1}\wedge \xi _{2}]=\hat{T}_{\xi _{1}+\xi _{2}}\ \exp
[\frac{-i}{\hbar }D_{3}(\xi _{1},\xi _{2})],  \label{eq:tt}
\end{equation}
where $D_{3}$ is the symplectic area of the triangle determined by two of
its sides. Evidently, the inverse of the unitary operator $\hat{T}_{\xi
}^{-1}=\hat{T}_{\xi }^{\dag }=\hat{T}_{-\xi }$ .

The set of operators corresponding to phase space reflections $\hat{R}_{x}$
about points $x=(p,q)$ in phase space, is formally defined in \cite{ozrep}
as the Fourier transform of the translation (or Heisenberg) operators 
\begin{equation}
\widehat{R}_{x}\equiv (4\pi \hbar )^{-L}\int d\xi \quad e^{\frac{i}{\hbar }%
x\wedge \xi }\widehat{T}_{\xi }.  \label{eq:rint}
\end{equation}
Their action on the coordinate and momentum bases are 
\begin{eqnarray}
\hat{R}_{x}\left| q_{a}\right\rangle  &=&e^{2i(q-q_{a})p/\hbar }\;\left|
2q-q_{a}\right\rangle   \label{rqpl} \\
\hat{R}_{x}\left| p_{a}\right\rangle  &=&e^{2i(p-p_{a})q/\hbar }\;\left|
2p-p_{a}\right\rangle ,
\end{eqnarray}
displaying the interpretation of these operators as reflections in phase
space. Also, Using the coordinate 
representation of the coherent state (\ref{eq:qX}) 
and the action of reflection on the coordinate basis (\ref{rqpl}), we can see
 that the action  of the reflection operator $\widehat{R}_{x}$ on a coherent state $\left|
X\right\rangle $ is the $x$ reflected coherent state 
\begin{equation}
\widehat{R}_{x}\left| X\right\rangle =\exp \left( \frac{i}{\hbar }X\wedge
x\right) \left| 2x-X\right\rangle .
\end{equation}

This family of operators have the property that they are a decomposition of
the unity (completeness relation) 
\begin{equation}
\hat{1}=\frac{1}{2\pi \hbar }\int dx\ \hat{R}_{x},  \label{1R}
\end{equation}
and also they are orthogonal in the sense that 
\begin{equation}
Tr\left[ \hat{R}_{x_{1}}\ \hat{R}_{x_{2}}\right] =2\pi \hbar \;\delta
(x_{2}-x_{1}).  \label{trR}
\end{equation}
Hence, an operator $\hat{A}$ can be decomposed in terms of reflection
operators as follows 
\begin{equation}
\hat{A}=\frac{1}{2\pi \hbar }\int dx\ A_{W}(x)\ \hat{R}_{x}.  \label{rep}
\end{equation}
With this decomposition, the operator $\hat{A}$ is mapped on a function $%
A_{W}(x)$ living in phase space, the so called Weyl-Wigner symbol of the
operator. Using (\ref{trR}) it is easy to show that $A_{W}(x)$ can be
obtained by performing the following trace operation 
\begin{equation*}
A_{W}(x)=Tr\left[ \hat{R}_{x}\ \hat{A}\right] .
\end{equation*}
Of course, as it is shown in \cite{ozrep}, the Weyl symbol also takes the
usual expression in terms of matrix elements of $\hat{A}$ in coordinate
representation 
\begin{equation*}
A_{W}(x)=\int \left\langle q-\frac{Q}{2}\right| \hat{A}\left| q+\frac{Q}{2}%
\right\rangle \exp \left( -\frac{i}{\hbar }pQ\right) dQ.
\end{equation*}

It was also shown in \cite{ozrep} that reflection and translation operators have the
following composition properties 
\begin{equation}
\widehat{R}_{x}\widehat{T}_{\xi }=\widehat{R}_{x-\xi /2}e^{-\frac{i}{\hbar }%
x\wedge \xi }\ ,  \label{eq:rt}
\end{equation}
\begin{equation}
\widehat{T}_{\xi }\widehat{R}_{x}=\widehat{R}_{x+\xi /2}e^{-\frac{i}{\hbar }%
x\wedge \xi }\ ,  \label{eq:tr}
\end{equation}
\begin{equation}
\widehat{R}_{x_{1}}\widehat{R}_{x_{2}}=\widehat{T}_{2(x_{2}-x_{1})}e^{\frac{i%
}{\hbar }2x_{1}\wedge x_{2}}  \label{eq:rr}
\end{equation}
so that 
\begin{equation}
\widehat{R}_{x}\widehat{R}_{x}=\widehat{1}\ .
\end{equation}
Now using (\ref{eq:rr}) and (\ref{eq:tr}) we can compose three reflections so that  
\begin{equation}
\widehat{R}_{x_{2}}\widehat{R}_{x}\widehat{R}_{x_{1}}=e^{\frac{i}{\hbar }%
\Delta _{3}(x_{2},x_{1},x)}\widehat{R}_{x_{2}-x+x_{1}}
\end{equation}
where $\Delta _{3}(x_{2},x_{1},x)=2(x_{2}-x)\wedge (x_{1}-x)$ is the area of
the oriented triangle whose sides are centered on the points $x_{2},x_{1}$
and $x$ respectively (see figure 1).

\begin{figure}[h]
\setlength{\unitlength}{1cm} 
\begin{picture}(0,16)(0,0)
%
\put(-8.0,8.0){\epsfxsize=8cm\epsfbox[0 0 550 550]{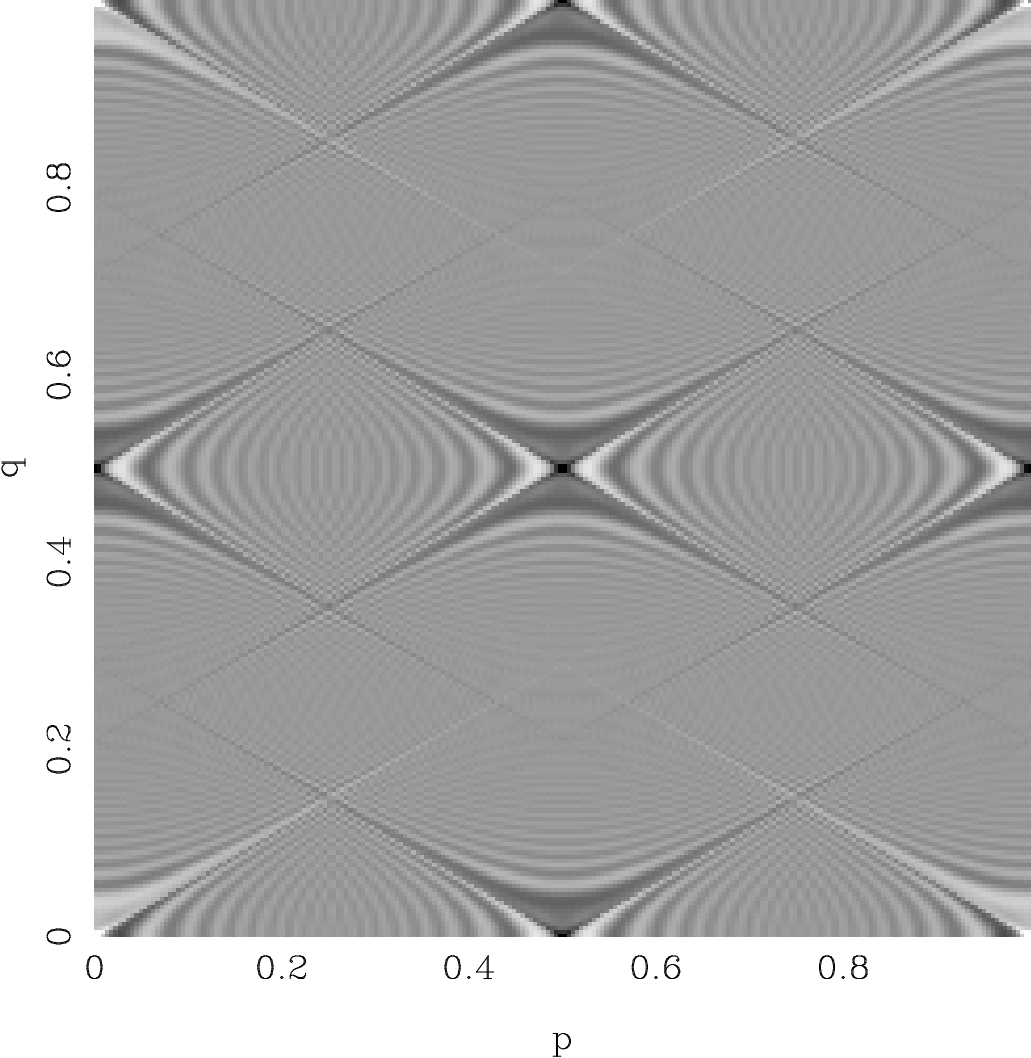}}
\put(0.0,8.0){\epsfxsize=8cm\epsfbox[0 0 550 550]{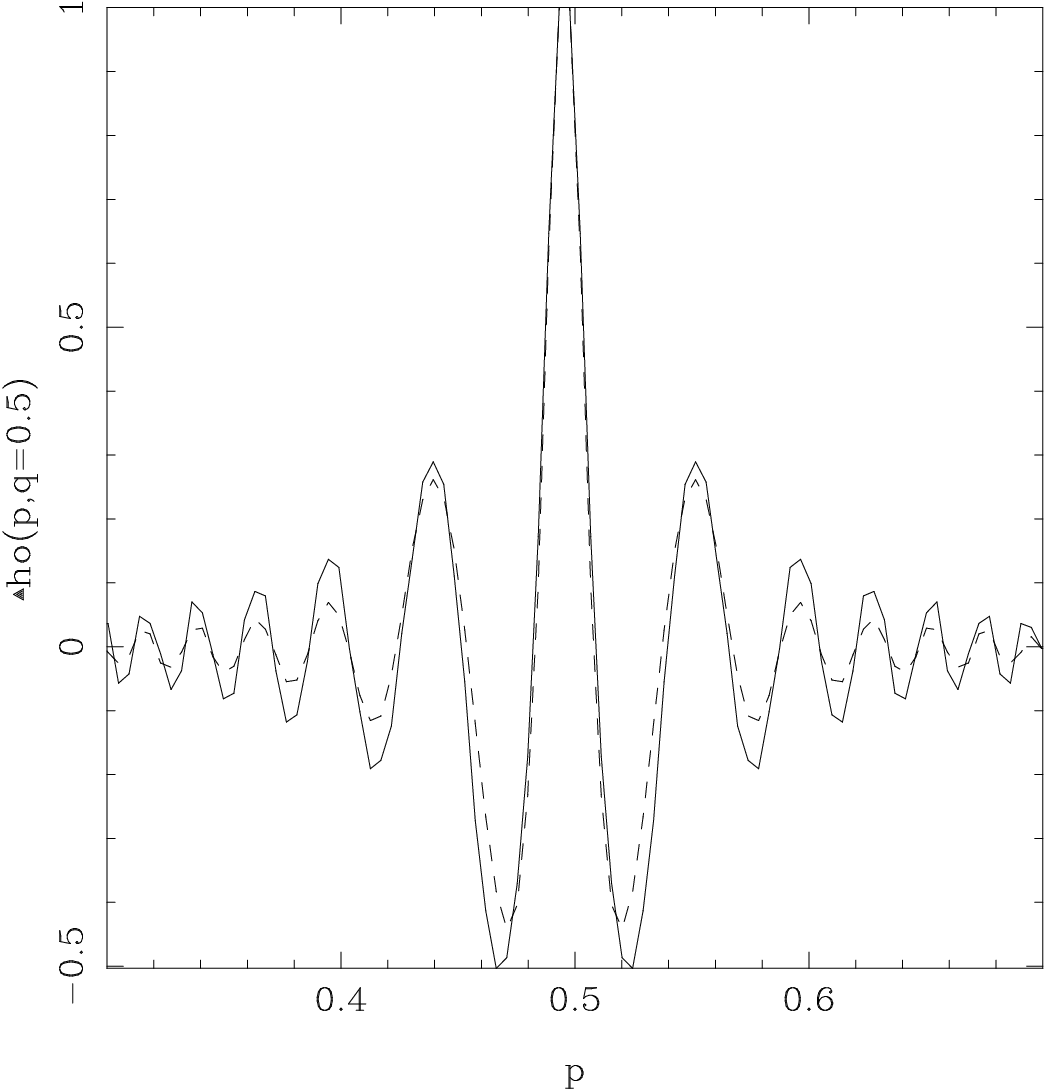}}
\put(-8.0,0.0){\epsfxsize=8cm\epsfbox[0 0 550 550]{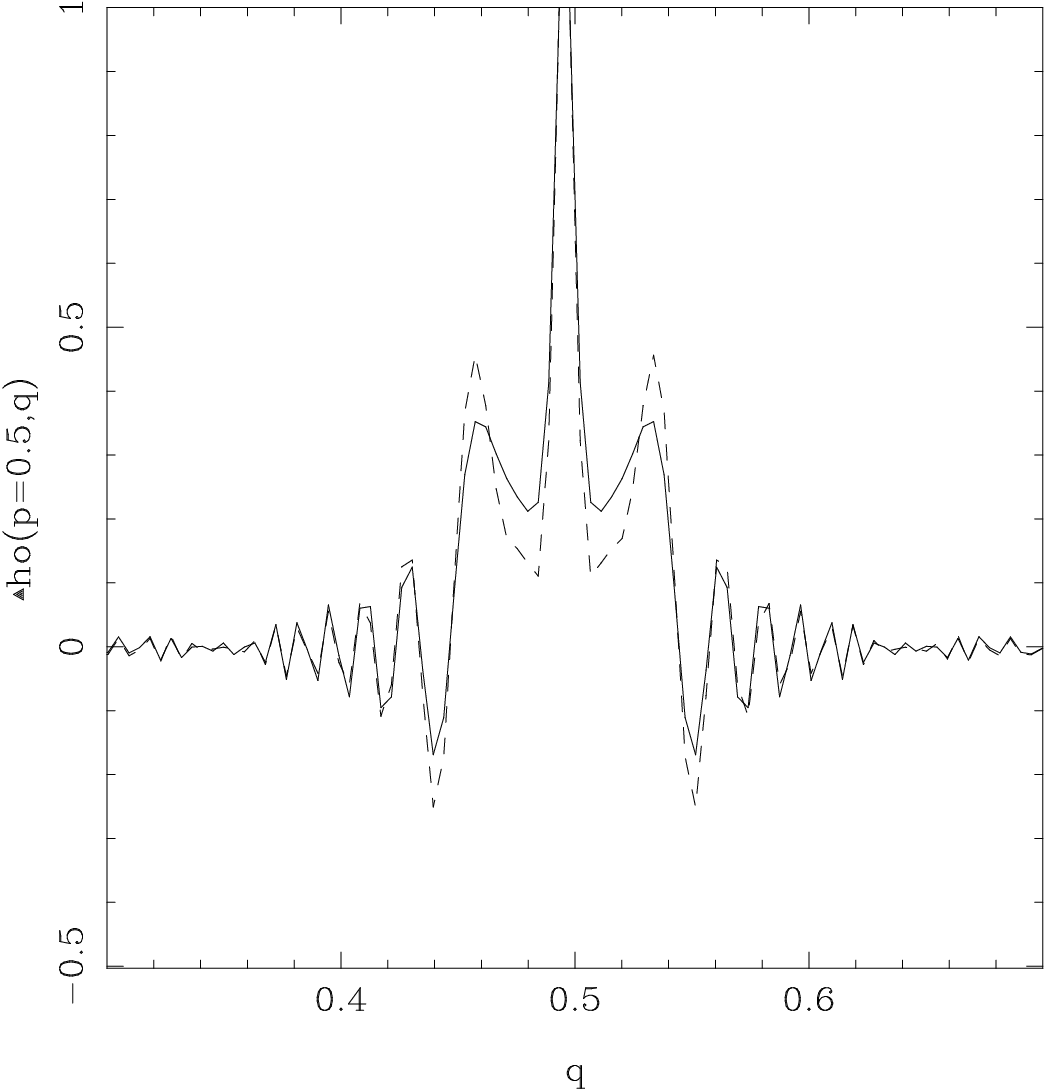}}
\put(0.0,0.0){\epsfxsize=8cm\epsfbox[0 0 550 550]{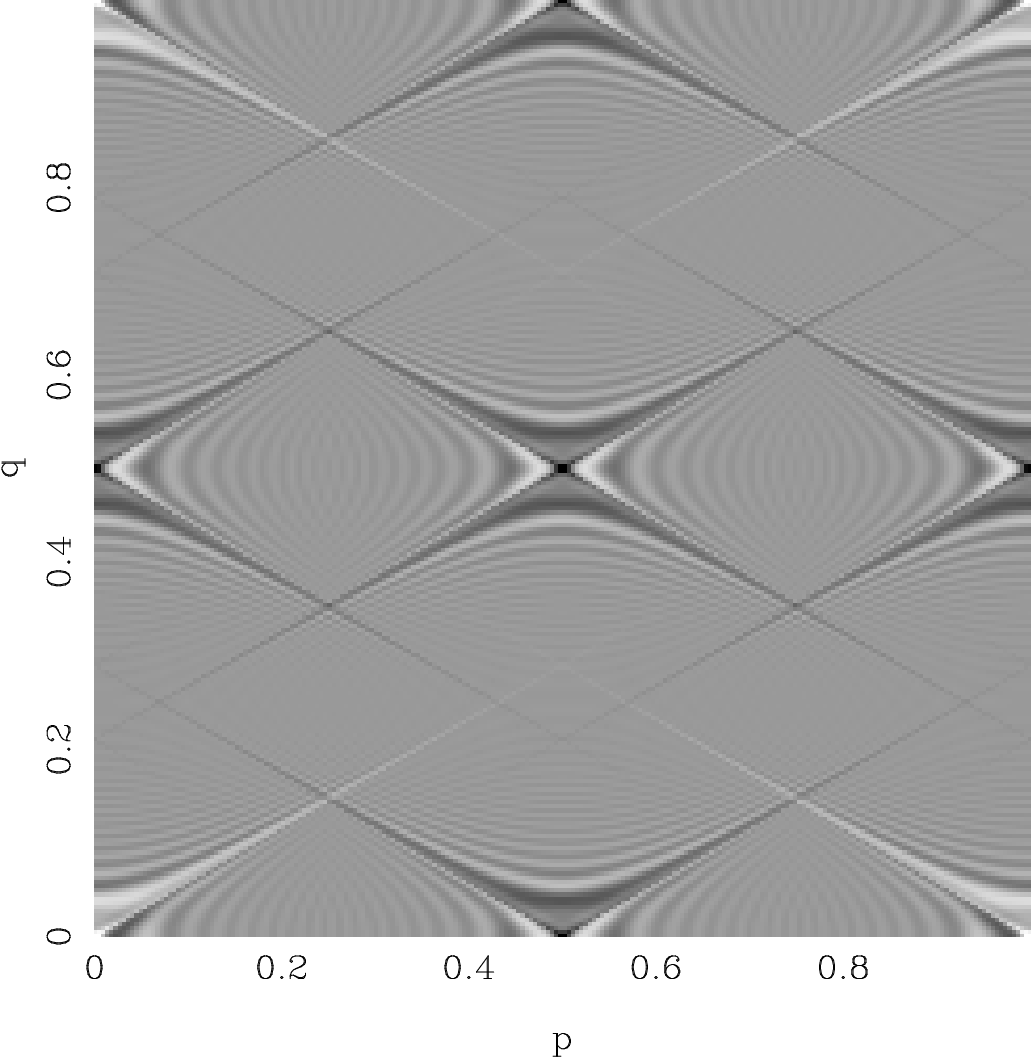}}
\par
\end{picture}
\vspace*{1.0pc}

\caption{\footnotesize  detail of the Scar Wigner  function with $N=223$ constructed on  the fixed point $(0,0)$ for a non Bohr quantized value of $\phi$.\\
(a) Exact result for the cat map.\\
(b) We compare sections of the exact and the semiclassical  Scar Wigner  functions near the periodic point for  $q=0.5$ (horizontal section). The solid line represents the section of the exact Scar Wigner function, while in  dashed lines we show the semiclassical approximation.  \\
(c) Idem (b) but for $p=0.5$ (vertical section).\\
(d) Semiclassical approximation $\mathbf{\rho}_{X,\phi}^{SC}(x)$.}
\label{fig.2}
\end{figure}

\end{document}